\documentclass[twocolumn,showpacs,preprintnumbers,amsmath,amssymb]{revtex4}
\usepackage{graphicx}
\usepackage{dcolumn}
\usepackage{bm}

\newcommand{\bef}{\begin{figure}}
\newcommand{\eef}{\end{figure}}

\newcommand{\be}{\begin{equation}}
\newcommand{\ee}{\end{equation}}
\newcommand{\bea}{\begin{eqnarray}}
\newcommand{\eea}{\end{eqnarray}}
\begin{document}

\title{Directed flow of open charm in Au+Au collisions at
  $\sqrt{s_{NN}}$ = 200~GeV using a quark coalescence model}

\author{Md Nasim$^{1}$ and Subhash Singha$^{2}$}
\affiliation{$^{1}$Indian Institute of Science Education and Research, Berhampur, India;\\
$^{2}$Kent State University, Ohio, USA}

\begin{abstract}
The directed flow ($v_{1}$) of open charm meson ($D^{0}$) is studied in Au+Au collisions at
$\sqrt{s_{NN}}$ = 200~GeV using A Multi-Phase Transport (AMPT) model
framework with partonic interactions (string melting version). Within this framework, it is found that
although the initial spatial eccentricity ($\epsilon_{1}$) of charm 
quark is smaller than light quarks, the charm quark $v_{1}$ magnitude
is found to be approximately 7 times larger than
that of the light u quark at large rapidity. This indicates that the charm
quarks can retain more information from initial condition than the
light quarks. We have studied the directed flow of $D^{0}$  as a
function of rapidity and transverse momentum using quark coalescence as the
mechanism for hadron production. Like charm quark, the $D^{0}$ $v_{1}$ magnitude is found to be about 7 times
larger than that of the light ($\pi$) hadrons at large rapidity.

\end{abstract}
\maketitle

\section{INTRODUCTION}

The main purpose of relativistic heavy-ion experiments is to understand the
formation and evolution of a strongly interacting matter, called Quark
Gluon Plasma (QGP)~\cite{qgp0}, which is expected to be formed micro-second after the
big bang. Experiments at the  Brookhaven Relativistic Heavy Ion Collider (RHIC) and at CERN Large Hadron
Collider (LHC) facilities established the existence of such strongly interacting
matter~\cite{whitepapers}, but the complexity in dynamics of the medium is still being
explored. Collective motion of the particles emitted from these collisions is
of special interest because it is sensitive to the equation of state of
the system. Directed flow ($v_{1}$) is characterized by the first harmonic
coefficient in the Fourier decomposition of
the momentum distribution of emitted particles ~\cite{flow_method, v1review},
\begin{equation}
v_{1}=\langle\cos(\phi-\Psi_{RP})\rangle,
\end{equation}
where $\phi$ denotes the azimuthal angle of emitted particles and
$\Psi_{RP}$ is the reaction plane subtended by the x-axis and impact
parameter direction. In this paper we consider the rapidity-odd
component of directed flow ($v_{1}^{\rm odd} (y) = -v_{1}^{\rm odd} (-y)$),
which refers to a sideward collective motion of emitted particles, and is a repulsive
collective deflection in the reaction plane. Whereas, the fluctuations in the
initial-state of the colliding nuclei can generate a rapidity-even
component of $v_{1}$ ($v_{1}^{\rm even} (y) = v_{1}^{\rm even} (-y)$) and it
is unrelated to the reaction plane\cite{v1_even}. In this paper $v_{1}$ denotes the
rapidity-odd component.

Model calculation~\cite{2rgamma} suggested that the directed flow near the beam
rapidity is initiated during the passage of two colliding nuclei. The
typical time scale of passing is $\sim 2R/ \gamma \sim 0.1$ fm/c for a
Au+Au collision at $\sqrt{s_{NN}}$ = 200~GeV, where $R$ and
$\gamma$ are the radius of nuclei and Lorentz factor respectively. So
the observable of directed flow is sensitive to the dynamics in the early stages of
nuclear collisions\cite{v1_early}. Both hydrodynamic~\cite{v1_hydro, Stoecker} and transport model~\cite{v1_transport} calculations have shown that the
directed flow at mid-rapidity, especially the baryons, are sensitive to
the equation of state of the system\cite{Rischke, Stoecker}. Several hydrodynamic calculations
suggested that the negative $v_{1}$-slope near mid-rapidity (called
``wiggle''~\cite{Stoecker, raimond} or
``anti-flow'' ~\cite{v1_anti}) could be a possible QGP signature\cite{Stoecker}. However, there are
the hadronic models with partial baryon-stopping and positive space
momentum correlations\cite{raimond}, and a hydro model full stopping with a tilted
source~\cite{tilt}  can also explain the anti-flow nature of $v_{1}$.
Recently, the STAR experiment at the RHIC has reported the measurements of directed flow
of several light hadron species ($\pi$, K, $K_{S}^{0}$, p,
$\Lambda$ and their anti-particles, and $\phi$) over the beam energy range 7.7--200~GeV\cite{STAR-BESv1, STAR-v1ncq}. 
Number of Constituent Quark (NCQ) scaling has been observed in
higher flow harmonics ($v_{2}$ and $v_{3}$) at both RHIC and LHC
energies~\cite{ncq1, ncq2, ncq-lhc}. Such scaling is interpreted as evidence of
quark degrees of freedom in the early stages of heavy-ion collisions. 
The recent $v_{1}$ measurements reported by STAR~\cite{STAR-v1ncq} found  to be consistent with the 
particles being formed via coalescence of constituent quarks. \\
The heavy quarks play a crucial role in probing the QGP medium,
because its mass is significantly larger than the typical temperature
achieved in such a collision. They are produced in hard partonic
scatterings during the early stages of collisions. The probability of
thermally produced heavy quarks are expected to be small in the high temperature
phase of QGP. Due to large mass, they decouple in the early stages of the collision.
The total number of charm quarks is frozen quite early in the history
of collision. So the heavy quarks are capable of retaining information of early time
dynamics. The measurement of directed flow of heavy quarks can offer
insight into the early time dynamics of the system. Apart from that,
recent measurements at the  RHIC\cite{STAR-HFv2} and LHC~\cite{ALICE-HFv2} have shown significant elliptic flow
for the charm hadrons. The flow magnitude of charm hadrons seems to
follow that of the light hadrons at mid-rapidity.  The $D^{0}$ $v_{2}$ from the AMPT
model~\cite{ampt} moderately explain recent STAR
data at mid-rapidity~\cite{nasim_hf_v2_review, nasim_cpod}. 

In this paper, we aim to study the directed flow of charm mesons
($D^{0} (\bar{u}c)$) in Au+Au collisions at 200~GeV within the
framework of AMPT model. Since the directed flow is generated in early times and also the charm
quark production limited to the primordial stage of the collisions,
the study of directed flow can offer insight into the initial dynamics
of the system. In this work, we have used string melting version (ver 2.26) of AMPT model~\cite{ampt}
(which includes parton coalescence) for the estimation of directed
flow. We have studied the $v_{1} (y, p_{T})$ of both heavy and light quarks. We have employed
dynamic coalescence mechanism to form hadrons from those quarks.
\begin{figure*}
\begin{center}
\includegraphics[scale=0.7]{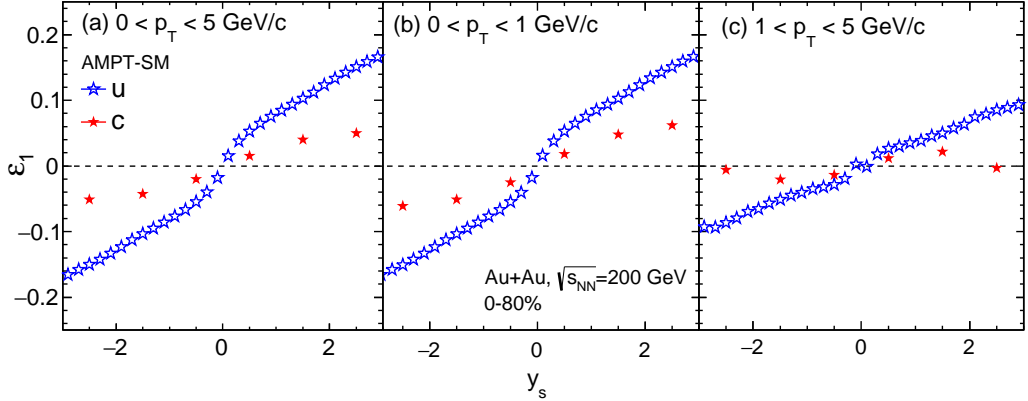}
\caption{(Color online) Initial geometric eccentricity
  ($\epsilon_{1}$) as a function of spatial rapidity ($y_{s}$) in Au+Au collisions at
  200 GeV  in three different $p_{T}$ regions( (a) 0--5~GeV/c, (b) 0--1~GeV/c and
  (c) 1--5~GeV/c )  for c and u quarks using AMPT-SM model.}
\label{fig:parton_ep1_rap}
\end{center}
\end{figure*} 

This paper is organized as follows. In the section \textrm{II}, we
discuss briefly AMPT model and dynamic coalescence of partons. 
Section \textrm{III} describes the directed flow $v_{1}$ of heavy and
light flavor mesons at 200~GeV Au+Au collisions using the AMPT framework.
(version 2.26). The section \textrm{IV} presents a summary of the
results.

\begin{figure}
\begin{center}
\includegraphics[scale=0.4]{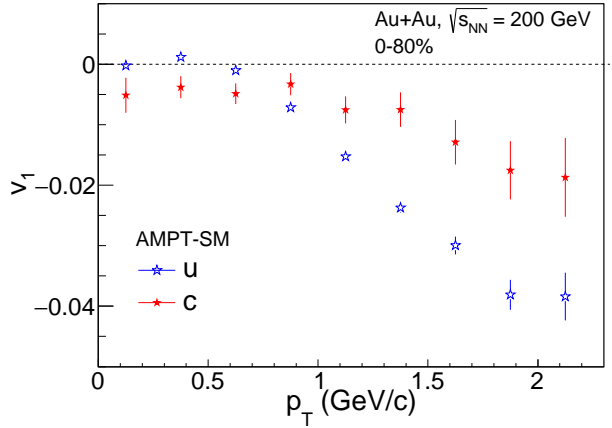}
\caption{(Color online) Comparison of $v_{1}$ for c and u quarks as a function of
  $p_{T}$ in positive rapidity region ($y > 0$) at 200~GeV Au+Au collisions using AMPT-SM model. }
\label{fig:parton_v1_pt}
\end{center}
\end{figure} 

\begin{figure*}[!ht]
\begin{center}
\includegraphics[scale=0.7]{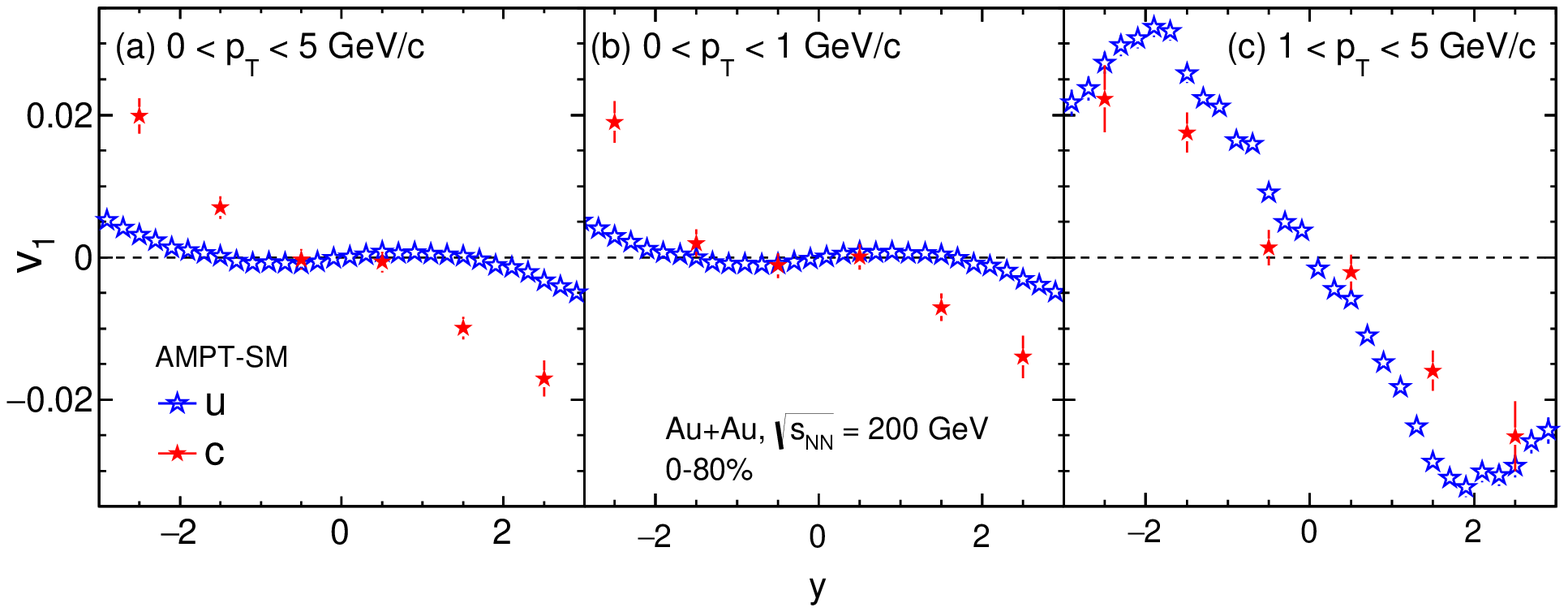}
\caption{(Color online) Comparison of $v_{1} (y)$ for c and u quarks
  in three different $p_{T}$ regions( (a) 0--5~GeV/c, (b) 0--1~GeV/c and
  (c) 1--5~GeV/c ) at 200~GeV Au+Au collisions using AMPT-SM model. }
\label{fig:parton_v1_rap}
\end{center}
\end{figure*} 

\section{The AMPT Model}
The AMPT is a hybrid transport model~\cite{ampt}. It uses the 
initial conditions from Heavy Ion Jet Interaction Generator (HIJING)~\cite{hijing}. However the minijet partons are made to undergo scattering before they are allowed
to fragment into hadrons.  The string melting (SM) version of the AMPT model 
(labeled here as AMPT-SM) is based on the idea that for energy densities beyond  a critical value of $\sim$ 1 GeV/$\rm {fm}^{3}$,
it is difficult to visualize the coexistence of strings (or hadrons) and partons. 
Hence the need to melt the strings to partons. 
Scattering among partons are modelled by Zhang's parton
cascade~\cite{ZPC}. Once the interactions stop, the partons then 
hadronizes through the mechanism of parton coalescence.
The parton-parton interaction cross section
in the string-melting version of the AMPT is given by
\begin{equation}
\sigma_{pp} = \frac{9 \pi \alpha_{S}^{2}}{2 \mu^{2}}
\end{equation}
For this study we set the strong coupling constant as $\alpha_{S}$ =
0.47 and the parton screening mass to be $\mu$ = 3.22 fm$^{-1}$. This leads to  $\sigma_{pp}$ = 3 mb. 
As the hadronization of heavy quarks is not implemented in AMPT-SM,
we use a dynamical coalescence model to form open charm mesons. Such a
model has been extensively used at both intermediate and high energies. In this model we use
phase-space information of partons at the freezeout to form open
charm mesons based on Wigner phase space function\cite{coal}. The
probability to form a meson from a pair of 
quark and anti-quark is given by,  
\begin{small}
\bea
\rho^W(\mathbf{r},\mathbf{k}) & = &  \int \psi \left( \mathbf{r}+\frac{\mathbf{R}}{2} \right) \psi^{\star}\left( \mathbf{r}-\frac{\mathbf{R}}{2} \right) \exp(-i \mathbf{k} \cdot \mathbf{R}) d^3 \mathbf{R}  \nonumber \\ 
& = & 8 \exp(-\frac{r^2}{\sigma^2}-\sigma^2 k^2)
\eea
\end{small}
where $R$ is the center-of-mass coordinate of the quarks or
anti-quarks and $\Psi$ is the quark wave function. The relative
momentum between the two quarks is 
 ${\bf {k}} = \frac{1}{m_{1}+m_{2}}(m_{2} \bf{p_{1}}$ - $m_{1} \bf{p_{2}})$.
Here $m_{i}$ is the mass of $i^{th}$ quark, and $p_{1}$ and $p_{2}$ are heavy
quark and light antiquark transverse momenta, respectively,
defined in the center-of-mass frame of produced meson~\cite{yong}.
For quarks, the Wigner phase-space densities are
obtained from the spherical harmonic oscillator wave functions,
\begin{equation}
\Psi({\bf{r_{1}}}, {\bf{r_{1}}}) = \frac{1}{(\pi \sigma^{2})^{3/4}}
\rm exp \bigg[\frac{-r^{2}}{2 \sigma^{2}}\bigg],
\end{equation}

where ${\bf {r}} = {\bf {r_{1}}} - {\bf{r_{2}}}$ and $\sigma$ is the
size parameter related to the root mean square radius as $\langle
r^{2} \rangle \ = (3/8)\sigma^{2}$~\cite{roli_d0_v2,greco_cmk,cmco}. In this paper, we have taken
$\langle r^{2} \rangle$ = 0.30 $fm^{2}$ for $D^{0}$  and $\langle r^{2} \rangle$ = 0.44 $fm^{2}$ for pion as predicted by the light-front
quark model~\cite{rms}.

\section{Results and Discussion}
The flow harmonic, $v_{1}$, quantifies the 1$^{st}$ order anisotropy
of particles of interest in the momentum space, and its magnitude is
a response of the initial anisotropy, the expansion dynamics and the
equation of state of the medium. Figure~\ref{fig:parton_ep1_rap}
presents the initial odd-eccentricity ($\epsilon_{1}$) of u and
c quarks as function of spatial rapidity ($y_{s}$) in Au+Au collisions at 200~GeV  in three different transverse momentum ($p_{T}$) regions. The $\epsilon_{1}$ can be extracted following the
equation~\cite{ampt_v1_junxu, liu_xu}: 

\begin{equation}
\epsilon_{1} = \langle \rm cos (\phi_{s} - \Psi_{RP}) \rangle,  
\end{equation}

where $\phi_{s}$ denotes the particle azimuthal angle, $\langle ...
\rangle$ denotes the average at a given rapidity and $\Psi_{RP}$ is
the reaction plane. In this paper, we have used the theoretical reaction plane
$\Psi_{RP}$=0 for the $v_{1}$. It is observed that the $\epsilon_{1}$
for c quarks is about 2--3 times smaller than that for the u
quarks in all $p_{T}$ regions. Next we try to see how this eccentricity is being transfered
to the directed flow. \\
The Figure~\ref{fig:parton_v1_pt} presents the $p_{T}$
differential $v_{1}$ for the c and u quarks in the forward rapidity
region. We observed that the u quark $v_{1}$ has a very strong $p_{T}$
dependence, while the c quark shows a weak dependence on $p_{T}$.
Figure~\ref{fig:parton_v1_rap} shows the rapidity dependence of c and
u quarks in three different $p_{T}$ regions. 
Motivation for showing $v_{1}$(y) in three different $p_{T}$ intervals
comes from $p_{T}$ dependence of $v_{1}$, as shown in Fig.~\ref{fig:parton_v1_pt}.
The panel (a) in Figure~\ref{fig:parton_v1_rap} presents $v_{1}(y)$ for $0 < p_{T} < 5$~GeV/c.
The magnitude of $v_{1}$ (first order anisotropy in momentum space)
for u quarks is few order smaller than the magnitude of $\epsilon_{1}$ (first order anisotropy
in coordinate space) with opposite sign in the forward and backward rapidities. Whereas for
the c quarks the magnitude of $\epsilon_{1}$ and $v_{1}$ is of similar
order. This is due the effect of system evolution in the partonic phase in the
AMPT model. All though the parton-parton interaction cross-section in the AMPT model is taken to be same (3mb) for all types of quarks, charm quark are less affected by the scattering due to its heavy mass. Therefore, change in  momentum (or $v_{1}$ = $<p_{x}>/<p_{T}>$) of charm quarks are less during the interaction with other light quarks.
We observe that full $p_{T}$ integrated $v_{1}$-values for c quarks ($~$0.02) is about 7
times larger than that of the u quark ($~$0.003) within the range $2.0 < |y| < 3.0$. This indicates that
the heavy c quarks retain more information about the initial
anisotropy than light u quarks, since initial $\epsilon_{1}$ of u quarks is larger than c quarks.  
However, we do not see any significant difference between $v_{1}$ of u and c quarks at mid-rapidity. Our model calculation suggested that rapidity dependence of flow harmonics of various identified hadrons need to measured in experiment to better understand the dynamics of the produced medium. In this paper, we have concentrated our calculation only on the $v_{1}$ co-efficient. 
The panel (b) and (c) presents the $v_{1}(y)$ in low ($0 < p_{T} < 1$~GeV/c) and higher
$p_{T}$ ($1 < p_{T} < 5$~GeV/c) regions. While at low $p_{T}$ the
magnitude of c quark $v_{1}$ is larger than the u quarks, at higher
$p_{T}$ their magnitudes are comparable. In AMPT-SM model, the $p_{T}$
integrated $v_{1}(y)$ of pions are actually dominated by the low $p_{T}$ ($< $1.0
GeV/c) pions due to a very sharp fall in pion $p_{T}$ spectra after $p_{T}$
= 1.0~GeV/c. We have observed that the $p_{T}$ spectra for charm
quarks are more harder than the light quarks. Although u-quarks has
large $v_{1}$ in the  range   $p_{T}$ $>$ 1.0~GeV/c, the  $p_{T}$
integrated $v_{1}(y)$ for $p_{T}$ $>$ 1.0~GeV/c
(Fig.~\ref{fig:parton_v1_rap}(c))  shows nearly same magnitude for
both ``charm'' and ``up'' quarks.\\
\begin{figure}[!h]
\begin{center}
\includegraphics[scale=0.4]{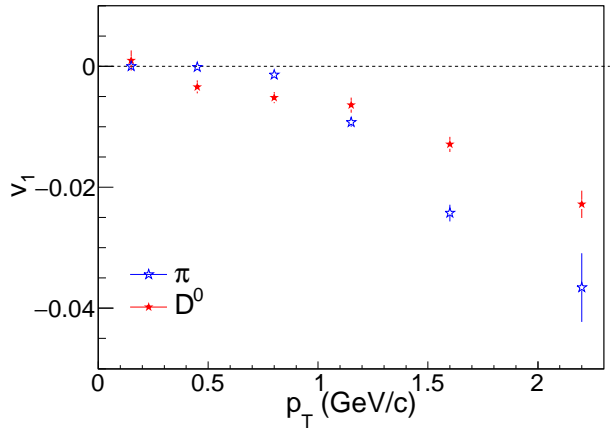}
\caption{(Color online) Comparison of $v_{1} (p_{T})$ in positive
  rapidity region ($y>0$) for $D^{0}$ and $\pi$ in 200~GeV Au+Au
  collisions using AMPT-SM model. }
\label{fig:D0_v1_pt}
\end{center}
\end{figure} 
Next, we employ dynamic coalescence mechanism, as described in
section~\textrm{II}, to form mesons from the  quarks at the freezeout. 
The u and $\bar{d}$ quarks are used to form pions, while c and
$\bar{u}$ quarks are used to get the $D^{0}$.
The Figure~\ref{fig:D0_v1_pt} presents
$p_{T}$ differential $v_{1}$ for $D^{0}$ and $\pi$'s in the forward
rapidity region ($y>0$). The $\pi$'s have a stronger $(p_{T})$
dependence of $v_{1} $ than for the $D^{0}$'s, which reflects the
similar behavior of the constituent quarks which is shown
in~Figure~\ref{fig:parton_v1_pt}. \\
\begin{figure*}[!ht]
\begin{center}
\includegraphics[scale=0.7]{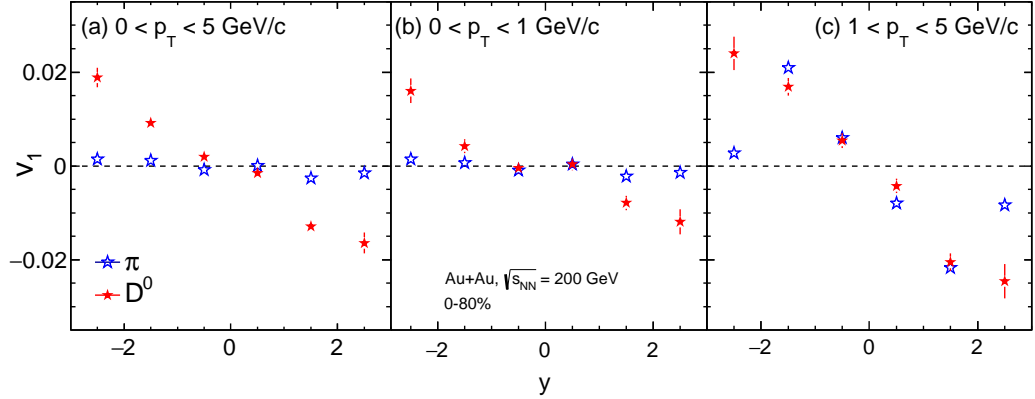}
\caption{(Color online) Comparison of $v_{1}(y)$ for for $D^{0}$ and
  $\pi$ in three different $p_{T}$ intervals ( (a) 0--5~GeV, (b) 0--1~GeV and
  (c) 1--5~GeV ) in 200~GeV Au+Au collisions using AMPT-SM model.}
\label{fig:D0_v1_rap}
\end{center}
\end{figure*} 
The Figure~\ref{fig:D0_v1_rap} shows the rapidity dependence of
$v_{1}$ for $D^{0}$ and $\pi$'s in three different $p_{T}$ intervals.  The
panel (a), (b) and (c) present $v_{1}(y)$ for $0 < p_{T} < 5$~GeV/c, $0 < p_{T} < 1$~GeV/c and $1 < p_{T} < 5$~GeV/c, respectively.
Fig.~\ref{fig:D0_v1_rap}(a) shows that the $D^{0}$ $v_{1}$ has large
magnitude than that of pions for $|y| > 1.0$.
The full $p_{T}$ integrated $D^{0}$ $v_{1}$  is found to be factor 7
times larger than that of pions within the range $2.0 < |y| < 3.0$.
The panel (b) and (c) represent similar observation as shows for partons in Fig.~\ref{fig:parton_v1_rap}.
Our observation from AMPT model calculation suggest that $D^{0}$
$v_{1}$ can be used as a useful probe, in addition to light hadrons
$v_{1}$, to study the initial state effect in heavy ion collisions.
There are recent hydro calculations~\cite{sandeep} that suggests that the $v_{1}$-slope of
heavy flavors can be sensitive probe of the initial matter distribution. The AMPT model with different dynamics for the charm
quarks hints towards the same direction. \\
A recent paper~\cite{santosh} predicted that the transient magnetic field in heavy-ion
collisions can induce a larger $v_{1}$ in heavy quarks than for light
quarks. Model also predicts opposite sign for charm and anti-charm
quarks due to the magnetic field. In future, one can study these effect
on charm $v_{1}$ within the AMPT model framework. We also look forward to the measurement of charm $v_{1}$ at both RHIC
and LHC energies.

\section{Summary and Conclusion}
In summary, we have presented the directed flow of heavy and light
flavor hadrons, and their constituent quark species in Au+Au
collisions at $\sqrt{s_{NN}}$=200~GeV using the string melting version
of AMPT model. Although the initial rapidity-odd eccentricity
($\epsilon_{1}$) in spatial coordinates for heavy
quarks are smaller than for the light quarks, the $v_{1}$ magnitude for heavy flavor hadrons is
approximately 7 times larger than that of the light hadrons at large rapidity. This is an interesting
observation, which tells us that the charm hadrons are capable of
retaining more information of the initial dynamics than the light
ones.  Any future  measurement of $D^{0}$ $v_{1}$ in a large rapidity window would be interesting to understand the initial dynamics in heavy-ion collisions. 

\noindent{\bf Acknowledgments}\\
Authors would like to thank Santosh Das, Sandeep Chatterjee,
Bedangadas Mohanty and Declan Keane for discussions and providing
fruitful suggestions. SS acknowledges financial support from DOE
project (grant DE-FG02-89ER40531), USA. Authors would like to
acknowledge hospitality at NISER-Jatni campus where a part of this
work has been done.


\normalsize
\begin{thebibliography}{99}

\bibitem{qgp0} J. C. Collins and M. J. Perry,
  Phys. Rev. Lett. {\bf{34}}, 1353 (1975);
S. A. Chin, Phys. Lett. {\bf{B 78}}, 552 (1978);
J. I. Kapusta, Nucl. Phys. {\bf{B 148}}, 461 (1979);
R. Anishetty, P. Koehler and L. D. McLerran, Phys. Rev.
{\bf{D 22}}, 2793 (1980).
\bibitem{whitepapers} I. Arsene  {\it et al.}  (BRAHMS Collaboration), Nucl. Phys. \textbf{A 757}, 1 (2005);
                      B. B. Back  {\it et al.} (PHOBOS Collaboration), Nucl. Phys. \textbf{A 757}, 28 (2005);
                      J. Adams  {\it et al.}   (STAR Collaboration),   Nucl. Phys. \textbf{A 757}, 102 (2005);
                      K. Adcox  {\it et al.}   (PHENIX Collaboration),
                      Nucl. Phys. \textbf{A 757}, 184 (2005).

\bibitem{flow_method} A. M. Poskanzer and S. A. Voloshin,  Phys. Rev. \textbf{C 58},
  1671  (1998).
\bibitem{v1review} S. Singha, P. Shanmuganathan and D. Keane,
  Adv. High Energy Phys. {\bf 2016}, 2836989 (2016).

\bibitem{v1_even} D. Teaney and L. Yan, Phys. Rev. {\bf C 83}, 064904
  (2011);
M. Luzum and J. Y. Ollitrault, Phys. Rev. Lett. {\bf 106}, 102301 (2011).
\bibitem{2rgamma} H. Sorge, Phys. Rev. Lett. \textbf{78}, 2309 (1997). 
\bibitem{v1_early} Y. Nara, A. Ohnishi, and H. St¨ocker,
  arXiv:1601.07692 [hep-ph];
V. P. Konchakovski, W. Cassing, Yu. B. Ivanov and V. D. Toneev,
Phys. Rev. {\bf C 90}, 014903 (2014);
\bibitem{v1_hydro} U. W. Heinz, in Relativistic Heavy Ion Physics, Landolt-Boernstein New Series, Vol. I/23, edited by R. Stock
(Springer Verlag, New York, 2010);
\bibitem{Rischke}
D. H. Rischke {\it et al.}, Heavy Ion Phys. {\bf 1}, 309 (1995).  
\bibitem{Stoecker}
H. Stocker, Nucl. Phys. A {\bf 750}, 121 (2005).
\bibitem{v1_transport} S. A. Bass et al., Prog. Part. Nucl. Phys. {\bf
    41}, 255 (1998);
M. Bleicher, E. Zabrodin, C. Spieles, S. A. Bass, C. Ernst, S. Soff,
L. Bravina, M. Belkacem, H. Weber, H. Stocker and W. Greiner, J. Phys. G 25, 1859 (1999).
\bibitem{raimond} R. J. M. Snellings, H. Sorge, S. A. Voloshin,
  F. Q. Wang and N. Xu, Phys. Rev. Lett. {\bf 84}, 2803 (2000).
\bibitem{v1_anti} J. Brachmann et al., Phys. Rev. {\bf C 61}, 024909 (2000).
\bibitem{tilt} P. Bozek and I. Wyskiel, Phys. Rev. {\bf C 81}, 2803 (2000).
\bibitem{STAR-BESv1}
L. Adamczyk {\it et al.} (STAR collaboration), Phys. Rev. Lett. {\bf
  112}, 162301 (2014).  
\bibitem{STAR-v1ncq} L. Adamczyk {\it et al.} (STAR Collaboration), arXiv:1708.07132
\bibitem{ncq1} 
J. Adams {\it et al.} ( STAR Collaboration), Phys. Rev.
Lett. {\bf 92}, 052302 (2004); B. Abelev {\it et al.} (STAR
Collaboration), Phys. Rev. C {\bf 75}, 054906 (2007); 
J. Adams {\it et al.} ( STAR Collaboration), Phys. Rev. C
{\bf 72}, 014904 (2005); B. I. Abelev {\it et al.} (STAR Collaboration),
Phys. Rev. Lett. {\bf 99}, 112301 (2007).
\bibitem{ncq2}
S. S. Adler {\it et al.} ( PHENIX Collaboration), Phys.
Rev. Lett.  {\bf 91}, 182301 (2003); S. Afanasiev {\it et al.}
(PHENIX Collaboration), Phys. Rev. Lett. {\bf 99},
052301 (2007); A. Adare {\it et al.} ( PHENIX Collaboration),
Phys. Rev. Lett. {\bf 98}, 162301 (2007); A. Adare
{\it et al.} ( PHENIX Collaboration), Phys. Rev. C {\bf 85},
064914 (2012).
\bibitem{ncq-lhc}
B. Abelev {\it et al.} (ALICE Collaboration), JHEP  {\bf 06}, 190
(2015); K. Aamodt {\it et al.} (ALICE Collaboration),
Phys. Rev. Lett. {\bf 105}, 252302 (2010).
\bibitem{STAR-HFv2}
L. Adamczyk {\it et al.} (STAR collaboration), Phys. Rev. Lett. {\bf
  118}, 212301 (2017).  
\bibitem{ALICE-HFv2}
B. Abelev {\it et al.} (ALICE collaboration), Phys. Rev. Lett. {\bf
  111} (2013) 102301
\bibitem{ampt} Zi-Wei Lin and C. M. Ko, 
               Phys. Rev. \textbf{C 65}, 034904 (2002);
               Zi-Wei Lin {\it et al.},
               Phys. Rev. \textbf{C 72}, 064901 (2005);
               Lie-Wen Chen {\it et al.}, 
               Phys. Lett. \textbf{B 605} 95 (2005).
\bibitem{nasim_hf_v2_review}
M. Nasim, R. Esha and H. Z. Huang,  Adv.High Energy Phys. 2016 (2016)
7140231.
\bibitem{nasim_cpod}
M. Nasim (for STAR Collaboration) arXiv:1801.04164.
\bibitem{hijing}X. N. Wang and M. Gyulassy, Phys. Rev. \textbf{D 44},
  3501 (1991).
\bibitem{ZPC} B. Zhang, Comput. Phys. Commun. \textbf{109}, 193
  (1998).
\bibitem{coal} K. P. Das and R. C. Hwa, Phys. Lett. \textbf{B 68},  (1977) 459; Erratum Phys. Lett. \textbf{B 73} (1978) 504;
D. Molnar and S. A. Voloshin, Phys. Rev. Lett. \textbf{91} (2003) 092301;
V. Greco, C.M. Ko and P. Levai,  Phys.Rev.  \textbf{C 68} (2003) 034904;  
B. Zhang, Lie-Wen Chen and C. M. Ko, Phys.Rev.  \textbf{C 72} (2005) 024906.
R. J. Fries {\it et al.} Ann. Rev. Nucl. Part. Sci. \textbf{ 58}, (2008)177.


\bibitem{yong} Y. Oh {\it et al.}  Phys. Rev.C 79, 044905 (2009).

\bibitem{roli_d0_v2}
R. Esha, M. Nasim and H. Z. Huang, J. Phys. {\bf G44} (2017) no.4,
045109; arXiv:1603.02700v2
\bibitem{greco_cmk} V. Greco, C.M. Ko and R. Rapp, Phys. Lett. \textbf{B  595} (2004) 202.
\bibitem{cmco}  L. W. Chen and C. M. Ko , Phys. Rev. \textbf{C 73}, (2006) 044903 .

\bibitem{rms}  C.-W. Hwang, Eur. Phys. J. C 23, 585 (2002).

\bibitem{ampt_v1_junxu}
C. Q. Guo, C. J. Zhang and Jun Xu, Eur.Phys.J. {\bf A53} (2017) no.12,
233 
\bibitem{liu_xu} H. Liu, S. Pantikin and N. Xu, Phys. Rev. \textbf{C
    59}, 348 (1999).
\bibitem{santosh}
S. Das, S. Plumari, S. Chatterjee, J. Alam, F. Scardina and V. Greco,
Phys. Lett. {\bf B 768}, 260-264 (2017).
\bibitem{sandeep}
S. Chatterjee, P. Bozek, arXiv:1712.01189.

\end{thebibliography}
\end{document}